\documentclass[12pt]{JHEP3}

\usepackage{amssymb}
\usepackage{graphics}
\usepackage{epsfig}
\usepackage{graphicx}
\usepackage{subfigure}
\usepackage{bm}

\def\beq{\begin{equation}}
\def\eeq{\end{equation}}
\def\baq{\begin{eqnarray}}
\def\eaq{\end{eqnarray}}

\title{Generation of the Higgs Condensate and Its Decay after Inflation}

\author{Kari Enqvist, Tuukka Meriniemi and Sami Nurmi
\\
University of Helsinki and Helsinki Institute of Physics, P.O. Box
64, FI-00014, Helsinki, Finland }

\abstract {We investigate the dynamics of the Standard Model higgs
with a minimal coupling to gravity during and after inflation. In
the regime where the Standard Model vacuum is stable, we find that
the higgs becomes a light spectator field after about 30 efolds of
inflation, irrespectively of its initial value. Once the higgs has
become light, its root-mean-square value $h_*$ relaxes to
equilibrium in about 85 efolds for the inflationary scale of
$H_*=10^{4}$ GeV and in 20 efolds for $H_*=10^{10}$ GeV. The
equilibrium value is given by $h_*\sim 0.36 \lambda_*^{-1/4}H_{*}$,
where $\lambda_*=0.09 ... 0.0005$ is the higgs self coupling at the
scales $H_*=10^{4} ... 10^{10}$ GeV. We show that the main decay
channel of the higgs condensate after inflation is the resonant
production of Standard Model gauge bosons. For a set of parameters
we find that a significant part of the condensate has decayed in
between 340 and 630 Hubble times after the onset of higgs
oscillations, depending on $H_*$ in a non-trivial way. The higgs
perturbations correspond to isocurvature modes during inflation but
they could generate significant adiabatic perturbations at a later
stage for example through a modulation of the reheating stage.
However, this requires that the inflaton(s) decay no later than a
few hundred Hubble times after the onset of higgs oscillations. }

\preprint{HIP-2013-09/TH}

\begin{document}

\section{Introduction}
The exact mechanism for inflation is not known, although a
consistent description is obtained by assuming a slowly rolling
scalar field, the inflaton. Since the discovery of the higgs at LHC
\cite{Aad:2012tfa}, we know that there exists at least one other
scalar field, which during inflation appears to be a spectator field
with no dynamical role. While the higgs could in principle also act
as the inflaton, this possibility is not generic but requires a
strong non-minimal coupling to gravity \cite{Bezrukov:2007ep} and
has also been argued to have unitarity problems
\cite{Burgess:2010zq}. Although the spectator fields are
insignificant during inflation, they may become important at a later
stage. For instance, they could contribute to the generation of the
primordial curvature perturbation, as in the curvaton scenario
\cite{curvaton}. Their post-inflationary dynamics could also be
interesting on its own right.

If the Standard Model higgs is a light field during inflation, it is
also subject to fluctuations with an almost scale invariant
spectrum. These fluctuations may contribute to metric perturbations
in several ways. For instance, the higgs could act as a curvaton
although the viability of this mechanism is hampered by the quartic
form of the potential \cite{Choi:2012cp}, however see also \cite{Kunimitsu:2012xx}.
Another possibility would be that the higgs is coupled to the
inflaton and generates perturbations by modulating the end of
inflation \cite{Lyth:2005qk}, or the reheating process
\cite{Dvali:2003em}. Indeed, it has been argued
\cite{DeSimone:2012qr} that the entire origin of the observed
primordial perturbations could be due to inflationary fluctuations
of the higgs modulating the reheating process. However, as the
result of the quartic self-interaction of the higgs field, the
perturbations generated by this mechanism typically have significant
non-Gaussian features. With the Planck limit of $f_{\rm NL}=2.7\pm
5.8$ \cite{Ade:2013ydc}, these constrain the range of feasible
inflationary scales from the below
\cite{DeSimone:2012gq,Choi:2012cp}.

Independently of these possibilities, it is of interest to find out
what actually happens to the higgs condensate generated by the
inflationary fluctuations. This is an issue that might have
consequences also for the electroweak phase transition in the early
universe.

While inflation lasts, the higgs generically is in slow roll, as we
will discuss below. After the end of inflation and onset of inflaton
oscillations, also the higgs field begins to oscillate in its
potential once its effective mass roughly equals the Hubble rate.
Because its energy density is subdominant, these oscillations take
place in the background determined by the inflaton dynamics. As the
higgs oscillates, it eventually starts to decay to other Standard
Model particles it couples to. The decay could be either
perturbative or non-perturbative. If the higgs field were to decay
before the inflaton, there would be no possibility for generating
perturbations through its modulation of the inflaton decay.

The dynamical evolution of the higgs field and its decay after
inflation is a general problem which we address in this paper by
analysing both the perturbative and non-perturtubative decay
processes. Requiring that the higgs decays after the inflaton, which
would be a necessary condition for it to source the generation of
primordial perturbations through the modulated reheating, turns out
to be a non-trivial constraint for two reasons: the higgs field
starts to oscillate relatively soon after the end of inflation, and
its couplings to other Standard Model fields are large.

The paper is organized as follows. In Sect. 2 we address the
dynamics of the higgs field first during inflation, when its
root-mean-square value is evolving towards the equilibrium, and then
discuss the motion of the condensate after inflation. In Sect. 3 we
study the perturbative decay of the higgs condensate, discussing
blocking factors and computing the appropriate decay widths. In
Sect. 4 we investigate the non-perturbative decay and show that the
main decay process of the higgs condensate is the resonant
production of weak gauge bosons. We estimate the time scale when the
backreaction kicks in and the fraction of energy transferred to the
decay products becomes significant. Sect. 5 summarizes the reults
and contains a discussion about the higgs modulating the inflaton
decay.

\section{Dynamics of the higgs field}
\subsection{During inflation}
We consider the dynamics of the Standard Model higgs assuming the
Standard Model is valid up to the inflationary energy scale,
characterized by the Hubble rate $H_*$. We also assume that any
non-minimal coupling between the higgs field and gravity can be
neglected. For large field values, $h\gg v=246$ GeV, the bare higgs
mass is negligible compared to the quartic self-interaction term and
the effective higgs potential is well approximated by
  \beq
  V(h)=\frac{\lambda(\mu)}{4} h^4\ .
  \label{eq:HiggsPot}
  \eeq
Here $\lambda$ is the running self-coupling and we set the
renormalization $\mu$ equal to the higgs amplitude during inflation,
$\mu \sim h_*$. This choice serves to minimize radiative corrections
of the form ${\rm ln}(h/\mu)$ to the effective higgs potential. The
dependence of the effective potential on $\mu$ is spurious and the
choice $\mu=h_*$ is simply made for computational convenience. As we
will discuss below, the typical higgs values during inflation lie
somewhat below the inflationary scale $H_{*}$, so that we also have
$\mu=h_*\lesssim H_{*}$.

With this framework, the running of the coupling $\lambda(\mu)$, and
thereby the higgs potential, is fully determined and has been
computed up to three-loop precision \cite{Degrassi:2012ry}. The
coupling decreases towards larger energies and eventually hits zero
$\lambda(\mu_{\rm inst})=0$ at an instability scale $\mu_{\rm
inst}$. For the best fit values of Standard Model parameters the
instability scale lies around $\mu_{\rm inst}\sim 10^{11}$ GeV and
is pushed up to $\mu_{\rm inst}\sim 10^{15}$ GeV by decreasing the
top mass to $m_t= 171.0$ GeV, $3$-$\sigma$ below the best fit value
\cite{Degrassi:2012ry}. At energies above the instability scale, the
coupling $\lambda$ becomes negative implying an instability of the
Standard Model vacuum. Here we require the higgs potential remains
stable during inflation $\lambda(\mu)>0$. This bounds the
inflationary scale $H_{*}$ and the higgs value $h_{*}$ to lie below
the instability scale $\mu_{{\rm inst}}$. The coupling is bounded by
$\lambda\lesssim 0.1$ for field values $h_{*}\gtrsim 10^2$ GeV and
by $\lambda\lesssim 0.01$ for $h_{*}\gtrsim 10^{8}$ GeV
\cite{Degrassi:2012ry}. The higgs mass at inflationary scales is
then $V''(H_*)\gg (125\,{\rm GeV})^2$, unless $h_{*}\simeq \mu_{\rm
inst}$, and ignoring the bare mass in (\ref{eq:HiggsPot}) is indeed
well justified.

While the higgs potential is fully determined the initial field
value at the onset of inflation is an a priori unknown parameter.
For large initial values $h\gtrsim H/\sqrt{\lambda}$ the field is
effectively massive and oscillates around the minimum of its
potential. The amplitude decreases exponentially during inflation,
$h\sim e^{-N}$. The field enters very fast the regime $h\ll
H/\sqrt{\lambda}$ where its effective mass is small compared to the
Hubble rate. Even if the higgs started with an initial value of
$h\sim M_P$ (which would imply a very low top mass to keep the
vacuum stable) this would take only about 32 (18) efolds for the
inflationary scale $H_*=10^{4}$ GeV ($H_*=10^{10}$ GeV). The higgs
field will then slowly evolve towards the minimum of its potential
following the classical slow roll dynamics until the classical drift
has decreased comparable to the source term from quantum
fluctuations at around $V'\sim H^3$. From this point on, the quantum
fluctuations of the higgs field stretched to superhorizon scales
effectively make the field amplitude to perform a random walk.

Using the stochastic approach
\cite{starobinsky&yokoyama}, one may obtain an equilibrium
probability distribution for the mean field value $h_*$ during
inflation. It is given by
  \beq\label{equlibrdist}
  P(h)=C\, {\rm exp}\,\left(-\frac{2\pi^2\lambda_* h_*^4}{3
  H_*^4}\right)\ ,
  \eeq
where $C$ is a normalization constant.

Strictly speaking, the equilibrium distribution is an asymptotical
state valid only after an infinity of efolds. Following the
evolution of the probability distribution from an initial state
peaked around some field value, one observes the distribution both
spreading out approaching the equilibrium form and its central value
moving towards the equilibrium result. For a quartic potential, the
spreading of the probability distribution is characterized by a
decoherence time (the time scale for the approach of the variance to
the equilibrium result) which in terms of efolds has been found
\cite{Enqvist:2012xn} to be given by $N_{\rm dec}\approx
6\lambda^{-1/2}$. For $\lambda\gtrsim 0.005$, which corresponds to
an inflationary scale of $H_*=10^{10}$ GeV, this yields $N_{\rm
dec}\approx 85$. For comparison, for the low inflationary scale of
$H_*=10^{4}$ GeV one would find $\lambda\gtrsim 0.09$ and
equilibration at $N_{\rm dec}\approx 20$. The relaxation time,
measuring the rate of approach of the central value of the
distribution towards the equilibrium result, is bigger by roughly a
factor of two \cite{Enqvist:2012xn}. Therefore, it appears quite
natural to assume that the higgs fluctuations on observable scales
follow the equilibrium distribution, provided that the horizon
crossing of the largest observable scales was preceded by a few tens
of extra e-foldings or more.

Making this relatively mild assumption, we may thus estimate the
typical field value and the effective higgs mass at the horizon
crossing of observable scales by the expectation values computed
from the distribution (\ref{equlibrdist}). For the higgs mass we
then find the result
  \beq
  \label{higgsmass}
  m_{h_*}^2 = 3\lambda\langle h_*^2\rangle \simeq 0.40 \lambda_*^{1/2}
  H_{*}^2\ ,
  \eeq
indicating that the higgs is a light field during inflating,
$\eta_{h}=V_{hh}/(3H^2)\lesssim {\cal O}(0.01)$.

The expectation value of the higgs amplitude over the entire
inflating patch is vanishing $\langle h_*\rangle=0$ as a result of
the symmetric potential. However, the variance is non-zero and the
typical higgs amplitude in a random patch of the size of our
observable universe is given by the root mean square value
  \beq
  \label{higgsvalue}
 h_*\sim \sqrt{\langle h_*^2\rangle}\simeq 0.36 \lambda_*^{-1/4}
 H_{*}\ .
  \eeq
Computing the corresponding energy density we find that the higgs
condensate constitutes only a tiny fraction of the total energy
density during inflation, $\rho_{h}/\rho_{\rm tot.}\sim 10^{-3}
(H_*/M_{\rm P})^2$. We reiterate that these results are generic
predictions for the Standard Model higgs assuming the Standard Model
remains as a valid description and its vacuum stable up to
inflationary energy scales.

\subsection{After inflation}

The actual mass of the higgs field in our observable patch during
inflation may of course deviate from the ensemble average
(\ref{higgsmass}) as a result of a statistical fluke. However,
regions where the higgs mass significantly differs from $m_{h_*}^2
={\cal O}(0.1)\lambda_*^{1/2} H_{*}^2$ after a few tens of e-folds
of inflation occur with suppressed probabilities. Using the
equilibrium distribution (\ref{equlibrdist}) we find that the
probability to obtain field values $h_{*}$ corresponding to
$\eta_{h}>0.1$ (i.e. $m_{h_*}^2
> 0.1 \times 3 H^2_*$) is as low as $10^{-4}$ for $\lambda_*=0.01$
and it decreases exponentially for larger effective higgs masses.
This simple line of reasoning suggests that the Standard Model higgs
generally is a light spectator field during inflation. The
inflationary stage therefore gives rise to a nearly scale invariant
perturbations of the higgs field.

Due to the tiny energy density of the higgs condensate, its
perturbations amount to isocurvature fluctuations during inflation
which have little impact on the adiabatic metric fluctuations under
this epoch. Depending on the couplings between the higgs field and
the beyond Standard Model physics that drives the inflationary
stage, the higgs perturbations can however be converted into
observable metric perturbations after the end of inflation. This
would happen if the expansion history is sensitive to the exact
value of the higgs condensate, and its slight variation on
superhorizon scales, as for example in the modulated reheating
scenario \cite{Dvali:2003em} and the curvaton scenario
\cite{curvaton}.

A necessary condition for both scenarios is that the inflaton decay
and reheating of the universe occurs before the decay of the higgs
condensate. The decay rate of higgs into other Standard Model fields
is fully determined by the measured Standard Model parameters. We
can therefore unambiguously compute the decay time of the higgs
condensate\footnote{We ignore the decay channels of the higgs into
the beyond Standard Model inflaton(s). This could imply non-trivial
constraints on the inflaton sector and we plan to address these in a
future work.}. If the inflaton field decays before this epoch the
Standard Model higgs with a minimal coupling to gravity could
generate significant primordial perturbations through a modulation
of the reheating stage. However, if the inflaton decay occurs at a
significantly later stage the higgs fluctuations cannot be converted
into adiabatic perturbations by such a mechanism.

To discuss the higgs dynamics after the end of inflation we assume
for definiteness that the inflaton field, or the scalar
parameterizing the adiabatic direction at this stage, ends up
oscillating in a quadratic potential
  \beq
  V(\phi)=\frac{1}{2}m_{\phi}^2\phi^2\ .
  \eeq
The universe is then effectively matter dominated $H\propto
a^{-3/2}$. In the following we will approximate the Hubble rate at
the end of inflation by its value at the horizon crossing of
observable scales, $H_{\rm end} = H_{*}$.

After the end of inflation the higgs dynamics is classical and the
field starts to roll down from $h_*$ towards the minimum of the
potential. Soon after the end of inflation the scaling of the higgs
amplitude is given by $h\propto a^{-1}$, neglecting the logarithmic
running of the coupling $\lambda(h)$. Using (\ref{higgsmass}) we
then find that higgs becomes effectively massive, $m_{h}^2\sim
H_{\rm osc}^2$, at
  \beq
  \label{Hosc}
  \frac{H_{\rm osc}}{H_{*}}\sim \frac{1}{4}\lambda_{*}^{3/4}\ ,
  \eeq
which corresponds to $t_{\rm osc} \sim 3 \lambda_{*}^{-3/4}
H_{*}^{-1}$ Hubble times after the end of inflation. Putting in
numbers, for $\lambda_*\gtrsim {\cal O}(0.01)$ this gives $t_{\rm
osc} \lesssim  {\cal O}(10^2) H_{*}^{-1}$, or equivalently
$n_{\phi}\lesssim {\cal O}(10)$ inflaton oscillations. The higgs
oscillations therefore start relatively soon after the end of
inflation. Even if the inflaton would decay before higgs, the higgs
value at the decay time then in general differs from the value at
the time of inflation $h(t_{\rm dec})\neq h_*$.

In the following we will systematically analyze the decay channels
of the oscillating Standard Model higgs condensate. Identifying the
dominant decay channels we will work out the time when it decays and
compare it to the number of inflation oscillations.

\section{Perturbative decay}

The Standard Model higgs could decay perturbatively into quarks,
leptons and gauge fields. However, it turns out that the decay
channels into weak gauge bosons $h\rightarrow WW, h\rightarrow ZZ$
and top quarks $h\rightarrow tt$ that could lead to an efficient
perturbative decay are kinematically blocked due to masses generated
by the large higgs amplitude. The kinematically allowed perturbative
decay channels on the other hand lead to a very inefficient decay.

\subsection{Higgs decay into weak gauge bosons}

The effective masses of the weak gauge bosons generated by the higgs
expectation value are given by
 \beq
 \label{BosonMass}
  m_{W}=\frac{gh}{2}\ ,\qquad m_{Z}=\sqrt{g^{2}+g'^{2}}\,\frac{h}{2}\, >
  \,m_{W}\ .
 \eeq
These channels are kinematically blocked if $m_{h}<2m_{W}$ which
corresponds to $\lambda(\mu) < g^2(\mu)/3$. Setting $\mu\sim h$ this
translates into $h\gtrsim 10^{2}$ GeV so that for these higgs values
there is no phase space available for the decays $h\rightarrow WW,
h\rightarrow ZZ$. Using equation (\ref{higgsmass}) we find that it
takes about
  \beq
  t\sim \lambda(H_*)^{-3/8}\left(\frac{H_*}{10^2{\rm
  GeV}}\right)^{3/2}H_{*}^{-1}
  \eeq
Hubble times after the end of inflation until the higgs amplitude
has decreased from the inflationary value (\ref{higgsvalue}) to the
threshold $h\sim 10^2$ GeV at which the perturbative decay into
gauge bosons becomes possible. For $H_{*}\gtrsim 10^5$ GeV this
corresponds to, $t\gtrsim 10^6 H_{*}^{-1}$ Hubble times, or
equivalently $n_{\phi}\gtrsim 10^5$ inflaton oscillations. The
perturbative higgs decay into the weak gauge bosons therefore
remains blocked for a long time after the end of inflation unless
the inflationary scale would be extremely low, $H_{*}\ll 10^5$ GeV.

\subsection{Higgs decay into fermions}

The masses of quarks and leptons are given by
 \beq
 \label{QuarkMass}
m_{i}(h)=\frac{y_{i}(h)}{\sqrt{2}}h ,
 \eeq
where $y_i(h)$ denote the Yukawa couplings at the energy scale set
by the higgs value $h$. For the Standard Model higgs the
perturbative decay into top quarks $h\rightarrow{\bar t} t$ is
kinematically blocked. All other fermionic decay channels are
allowed unless the higgs value $h$ is tuned extremely close to the
instability scale $\lambda(h_{\rm inst})=0$.

To estimate the time scale of the higgs decay into fermions it
suffices to consider decays into bottom quarks. All other
kinematically allowed fermionic decay channels are suppressed by the
small Yukawas and their impact can be neglected here. The decay rate
of the process $h\rightarrow {\bar b}b$ for the higgs value $h_{\rm
osc}$ corresponding the onset of oscillations is given by
\cite{Djouadi}
  \beq
  \label{gammah2b}
  \Gamma(h\rightarrow bb)=\frac{3\sqrt{3\lambda}y_{b}^{2}h_{\rm osc}}{16\pi}\left(1-\frac{2y_{b}^{2}}{3\lambda}\right)^{3/2}
  \sim 10^{-6}\lambda_{*}^{3/4} H_{*} \ .
  \eeq
In the last step we have used that $y_{b}={\cal O}(10^{-2})$
\cite{Degrassi:2012ry} and $h_{\rm osc}\sim 0.1\lambda_{*}^{1/4}H_{*}$
which follows from equations (\ref{higgsmass}) and (\ref{Hosc}).
Setting $H=\Gamma(h\rightarrow bb)$ we find that it takes $t\gg 10^6
H_{*}^{-1}$ Hubble times, or $n_{\phi}\gg 10^5$ inflaton
oscillations, until the higgs can decay perturbatively to bottom
quarks. Decays into lighter quarks and fermions would occur even
later due to the smaller couplings.

\subsection{Higgs decay into photons}

The Standard Model higgs also decays into photons through loop
mediated processes. For $h\rightarrow \gamma\gamma$ the dominant
contributions come from the loops containing W-bosons and top
quarks. As we have noted above, these have masses much higher than
higgs, $m_{t}, m_{W}\gg m_{h}$, during the epoch we are considering.
In this limit, the decay rate of the higgs into two photons at the
onset of oscillations is given by \cite{Djouadi}
  \beq
  \label{gammah2gamma}
\Gamma\left(h\rightarrow\gamma\gamma\right)\sim0.02\alpha^{2}\lambda_{*}^{3/2}h_{\rm
osc}\ .
   \eeq
The fine structure constant $\alpha=g^2g'^2(4\pi(g^2+g'^2))^{-1}$ is
small, $\alpha\lesssim 0.02$, in the entire regime where the
Standard Model vacuum is stable \cite{Degrassi:2012ry}. From
equations (\ref{higgsvalue}) and (\ref{Hosc}) we obtain $h_{\rm osc
}\sim 0.1\lambda_{*}^{1/4}H_{*}$. Comparing equations
(\ref{gammah2gamma}) and (\ref{gammah2b}), we then immediately find
that
$\Gamma\left(h\rightarrow\gamma\gamma\right)\ll\Gamma\left(h\rightarrow
bb\right)$. Hence the decay channels into photons are always
insignificant compared to the higgs decay into bottom quarks .

The same holds true for the higgs decay $h\rightarrow\gamma Z$
mediated by quark and W-boson loops. This decay channel becomes
kinematically allowed for $h\lesssim10^{6}$ GeV as $m_{h}<m_{Z}$
\cite{Degrassi:2012ry}. The decay rate is approximately given by
\cite{Djouadi}
  \beq
\Gamma\left(h\rightarrow
Z\gamma\right)\sim10^{-5}g^{2}\lambda^{3/2}h_{\rm osc}\ .
  \eeq
Using that $h\leq h_{\rm osc}\sim 0.1\lambda_{*}^{1/4}H_{*}$ as the
decay channel opens and comparing with (\ref{gammah2b}) we find that
$\Gamma\left(h\rightarrow Z\gamma\right)\ll\Gamma\left(h\rightarrow
bb\right)$. These decay channels are therefore also insignificant.

\subsection{Higgs decay into gluons}

The higgs decay to two gluons is mediated by quark loops. The
largest contribution comes from the top quarks. The decay rate at
the onset of higgs oscillations is given by \cite{Djouadi}
  \beq
\Gamma\left(h\rightarrow gg\right)\sim10^{-5}
g_{s}^{4}\lambda^{3/2}h_{\rm osc}\ .
   \eeq
The factor $g_{s}^4$ comes from the two verteices which connects
gluons to a top quark loop. Using that $h_{\rm osc}\sim
0.1\lambda_{*}^{1/4}H_{*}$ and comparing with (\ref{gammah2b}) we
clearly see that $\Gamma\left(h\rightarrow
gg\right)\ll\Gamma\left(h\rightarrow bb\right)$ and the decay into
gluons is irrelevant compared to the decay channel into bottom
quarks.

\section{Non-perturbative decay}

In the previous section we have demonstrated that perturbative decay
of the Standard Model higgs becomes efficient only after a long period of Hubble times. This is due to the kinematical
blocking of the decay channels $h\rightarrow WW,ZZ,tt$ with the largest couplings.

The kinematical blocking is however instantaneously removed each
time when the oscillating higgs field crosses zero and the masses
induced by the higgs amplitude vanish. During these short time
intervals the Standard Model higgs can decay non-perturbatively into
gauge bosons and fermions. The non-perturbative decay of a generic
oscillating scalar field after inflation has been extensively
studied in the literature, see e.g.
\cite{Kofman:1997yn,StructureOfResonance,PreheatingOfFermions}. The
non-perturbative decay occurs much faster than the perturbative
decay and it turns out to be the dominant decay channel of the higgs
condensate, similar to what has been found in the context of higgs
inflation \cite{GarciaBellido:2008ab}.

Irrespectively of the type of decay products, the higgs fluctuations
cannot source the generation of adiabatic metric perturbations
through a modulation of the reheating stage after the higgs
condensate has decayed. Therefore, one of our goals here is to
identify the time scale of the non-perturbative higgs decay which
places a lower limit for the inflaton decay rate in a successful
modulated reheating scenario with the Standard Model higgs.

We will investigate both the bosonic and fermionic decay channels,
including the decay of the higgs condensate into higgs particles. In
our analysis we will neglect the backreaction of the resonantly
produced particles to the higgs dynamics, which eventually shuts
down the resonance. We will also neglect the perturbative decays of
the produced particles into other Standard Model fields. This is
justified as the perturbative decay widths are small compared to the
Hubble rate. For example, the decay width of the W-bosons to leptons
and quarks at the onset of higgs oscillations is given by
  \beq
  \frac{\Gamma_{W}^{\rm osc}}{H_{\rm osc}}\sim \frac{0.06 g^2 m_W}{H_{\rm
  osc}}\sim 0.01 \left(\frac{g}{0.5}\right)^3 \left(\frac{0.01}{\lambda}\right)^{1/2}\ .
  \eeq
We thus find that $\Gamma_W^{\rm osc} \ll H_{\rm osc}$ unless the
higgs value during inflation would be extremely close to the
instability point such that $\lambda\simeq 0$. The perturbative
decay of the resonantly produced W-bosons can therefore be neglected
at the beginning of the non-perturbative higgs decay. Similar
arguments hold for Z-bosons and top quarks produced by the resonant
higgs decay. Note that this is different from the decay of the
non-minimally coupled higgs field after the end of higgs inflation.
In that case the large higgs amplitude renders the perturbative
decay channels of the higgs decay products important which affects
the initial stages of the non-perturbative decay
\cite{GarciaBellido:2008ab}.

\subsection{Resonant production of gauge bosons and higgs particles}

We start by considering the non-perturbative production of weak
gauge bosons by the resonant decay of the higgs condensate. For
simplicity we will neglect the non-Abelian self-couplings of the
gauge bosons. This should not generate significant errors during the
first stages of the resonance when the number densities are still
small.

In the unitarity gauge the equation of motion for the transverse
components of the rescaled
$\mathcal{W}_{\mu}^{\pm}=a^{3/2}W_{\mu}^{\pm}$ gauge bosons is given
by
\beq
\label{Aeom}
\ddot{\mathcal{W}}_{\mu}^{\pm}(z,k)+\omega_{k}^{2}\mathcal{W}_{\mu}^{\pm}(z,k)=0\,,
\qquad \omega_{k}^{2}=\frac{k^2}{a^2\lambda h_{\rm
osc}^2}+q_{W}\frac{h(z)^2}{h_{\rm osc}^2}+\Delta\ .
\eeq
Here the resonance parameter is defined as
$q_{W}=m_{W}(x)^2/\left(\lambda h(x)^2\right)=g^2/\left(4
\lambda\right)$ and
$\Delta=-(3/4)({\dot{a}}/{a})^{2}-(3/2){\ddot{a}}/{a}$. For the
matter dominated background we have $\Delta=0$. Here use a re-scaled
cosmic time variable $z=\sqrt{\lambda h_{\rm osc}^2}(t-t_{\rm osc})$
and the over dot denotes a derivative with respect to $z$.

The equation of motion for the $Z$ bosons takes the same form
(\ref{Aeom}) but the resonance parameter $q_{W}$ gets replaced by
$q_{Z}=\left(g^2+g'{}^2\right)/\left(4 \lambda\right)$. The decay of
the higgs condensate into higgs particles is also controlled by an
equation of the same form but with $q_{W}$ replaced by
$q_{h}=m_{h}^2/\left(\lambda h^2\right)=3$.

Neglecting the backreaction of the resonantly generated particles,
the dynamics of the higgs condensate is determined by
\beq
\label{higgseom}
\frac{\ddot{h}}{h_{{\rm osc}}}+3\frac{H}{\sqrt{\lambda}h_{{\rm
osc}}}\frac{\dot{h}}{h_{{\rm osc}}}+\left(\frac{h}{h_{{\rm
osc}}}\right)^{3}=0\ ,
\eeq
where we have neglected the subdominant bare mass term of the higgs field, $\sqrt{2\lambda}v\propto v$. According to Eqs.~(\ref{higgsmass}) and (\ref{Hosc}) we have $h_{\rm osc}\sim 0.1\lambda_{*}^{1/4}H_{*}$. For $H_{*}\gtrsim10^{6}$ GeV we obtain $h_{\rm osc}\gg v=246$ GeV and therefore the effect of the bare mass would be insignificant. For $H_{*}\simeq10^{4}$ GeV we have $h_{\rm osc}/v\sim \mathcal{O}(1)$ so in this case the bare mass could be of some importance. Nevertheless, we ignore the bare mass term here because other effects, such as the thermal mass of the higgs due to the produced SM particles \cite{ThermalMass}, could be equally important. We leave elaborate consideration of these effects for later publications.

The induced gauge boson masses $m_{W} = g h/2 $,
$m_{Z}=\sqrt{g^2+g'{}^2} h/2$ vanish each time when the oscillating
higgs field crosses zero. The same also happens for the dominant
contribution to the higgs mass, $3\lambda h^2$, although the field
does not become strictly massless due to the small bare mass term.
During these short time intervals the effective masses change
non-adiabatically $|{\dot m_{W,Z,h}}|/m_{W,Z,h}^2\gtrsim 1$ and a
copious production of $W$, $Z$ and $h$ quanta can occur
\cite{Kofman:1997yn,StructureOfResonance}.

\begin{figure}[h!]
 \centering
    \includegraphics[height= 7 cm]{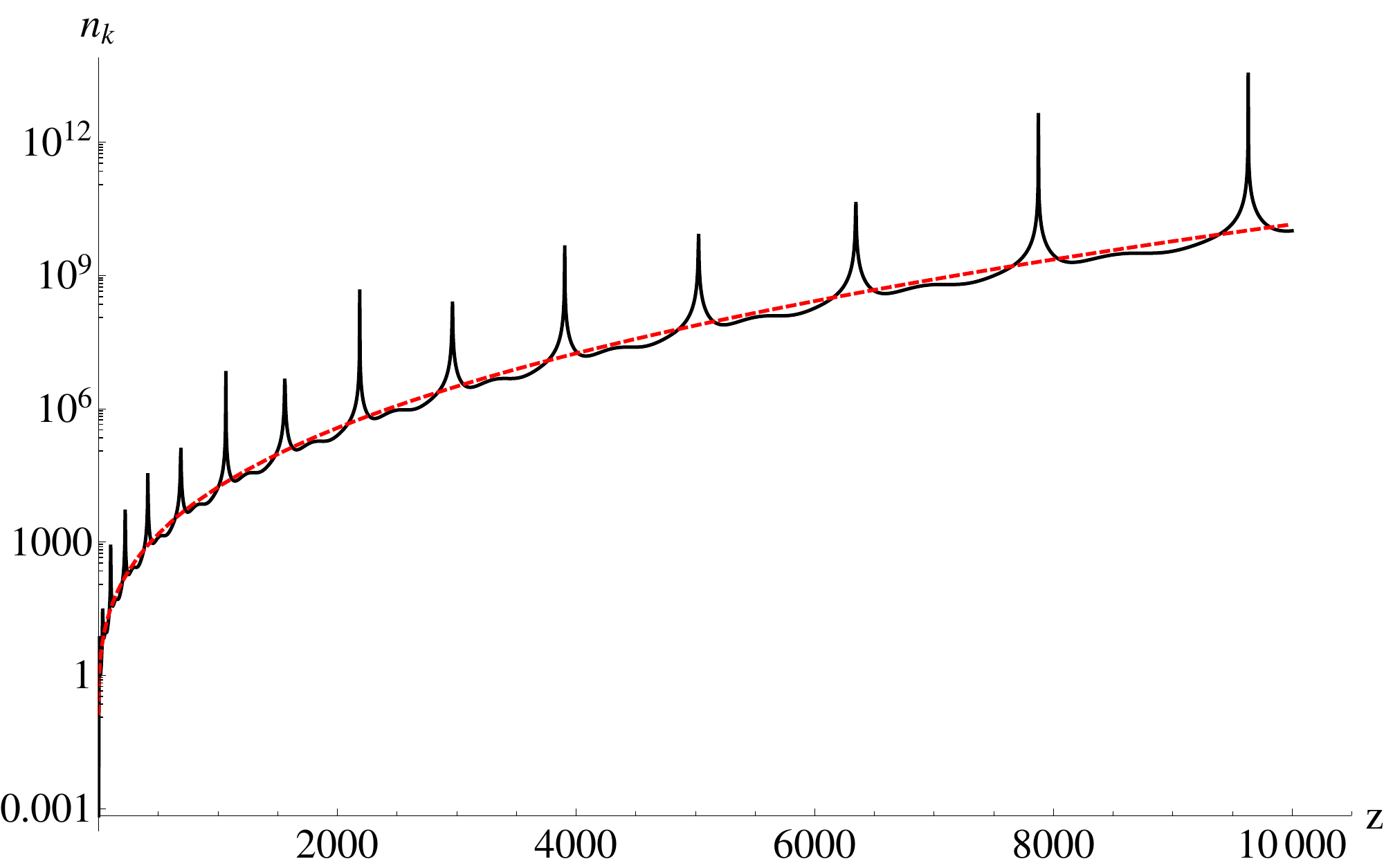}
    \caption{The evolution of the number density $n_k$ with $k=0$ and $q=1.5$ (black solid line) and exponential fitting (red dashed line) $n_{k}\propto\exp\left(4.6\mu_{k}z^{1/3}\right)$, with $\mu_k \approx 0.26$.
   }
    \label{fig:nk}
\end{figure}

The generation of the weak gauge bosons under the first stages of
the resonance can be straightforwardly analysed by solving the
equations of motion (\ref{Aeom}) and (\ref{higgseom}) numerically.
From the mode functions we then obtain the corresponding number
densities of the gauge bosons by using the relation
\cite{StructureOfResonance}
  \beq
  n_{k}=\frac{\omega_{k}}{2}\left(\frac{|\dot{\mathcal{W}}_{\mu}^{\pm}|^{2}}{\omega_{k}^{2}}+\left|\mathcal{W}_{\mu}^{\pm}\right|^{2}\right)-\frac{1}{2}\ .
  \eeq
The behaviour of the number density of $W$ or $Z$ bosons with
$q=1.5$ and $k=0$ is illustrated in Fig.~(\ref{fig:nk}). The growth
of the number density is approximatively exponential with
\footnote{In the conformal time, $x=\sqrt{\lambda h_{\rm
osc}^2}(\eta-\eta_{\rm osc})$, the number density evolves as
$n_{k}\propto \exp\left(2\mu_{k}x\right)$
\cite{StructureOfResonance}. }
  \beq
  \label{mu}
  n_{k}\propto \exp\left(4.6\mu_{k}z^{1/3}\right),
  \eeq
where $\mu_{k}$ is a characteristic exponent. In Table
(\ref{table:mu}) we show the numerically calculated characteristic
exponents $\mu_k$ of zero-momentum quanta for four different choices
of the inflationary scale, $H_*/{\rm
GeV}=10^4,\:10^6,\:10^8,\:10^{10}$, and both for the $W$ and $Z$
bosons. The inflationary scale $H_*$ sets the initial higgs value
through equation (\ref{higgsvalue}) and thereby eventually determines
the energy scale at which the Standard Model couplings are
evaluated.

The strength of the resonance is controlled by the resonance
parameter $q=q_{W,Z,h}$ in equation (\ref{Aeom}). The resonance is
typically stronger for larger values of $q$ but the relation is not
monotonic and for some values the resonance will not take place at
all. The structure of the resonance is closely related to that
encountered  in the conformally invariant scalar theory
$V=\lambda\phi^4/4+g^2\phi^2\chi^2/2$ \cite{StructureOfResonance}.
Indeed, apart from the small bare mass term which we have neglected and the matter
dominated background in our case, the gauge field equation of motion
(\ref{Aeom}) coincides with the conformally invariant form. For the
conformal model, the zero-momentum quanta are resonantly amplified
in the instability bands $q=n(n+1)/2...(n+1)(n+2)/2$, where $n$ is
an odd integer \cite{StructureOfResonance}. Thus, the first three
instability bands are $q=1...3$, $q=6...10$ and $q=15...21$. These
instability bands also seem to describe relatively well the
resonance in our case. This can also be seen in Table
(\ref{table:mu}) where the characteristic exponent $\mu_k$ is
non-zero inside the instability bands and outside the bands it
vanishes.

Usually quanta with low momenta are most efficiently produced and if
particles with $k=0$ are not excited then the resonance is
relatively weak. Thus, we consider here only the resonance bands
corresponding to $k=0$. As the resonance parameters $q_{W}$ and
$q_{Z}$ have a different dependence on the Standard Model couplings
we find that for most of the parameter space at least one of the
resonant decay channels into weak gauge bosons is efficient. It is
however also possible that both resonance parameters happen to be
outside the resonance bands. In such a situation the decay of the
higgs condensate would be delayed.

For the decay of the higgs condensate into higgs particles the
resonance parameter takes the value $q_h=3$. This lies on the edge of
the instability band \cite{StructureOfResonance} and therefore we
obtain an anomalously low characteristic exponent $\mu_k\approx
0.033$ for $k=0$. We thus find that the decay of the higgs
condensate into higgs particles is generically subdominant
compared to the decay into gauge bosons.

\begin{table}[ht]
\caption{Numerical values of the characteristic exponent $\mu_k$ of
$k=0$ modes for a set of different values of $H_*$.} \centering
\begin{tabular}{c c c c}
\hline\hline
$H_*$/GeV & $\lambda$ & $\left(q_{W},\:\mu_k\right)$ & $\left(q_{Z},\:\mu_k\right)$ \\ [0.5ex]
\hline
$10^4$ & $0.09$ & $\left(1.1,\:0.14\right)$ & $\left(1.5,\:0.26\right)$ \\
$10^6$ & $0.04$ & $\left(2.3,\:0.25\right)$ & $\left(3.2,\:0.00\right)$ \\
$10^8$ & $0.02$ & $\left(4.4,\:0.00\right)$ & $\left(6.2,\:0.14\right)$ \\
$10^{10}$ & $0.005$ & $\left(16,\:0.22\right)$ & $\left(24,\:0.00\right)$ \\ [1ex]
\hline
\end{tabular}
\label{table:mu}
\end{table}

\subsection{Time scale of the higgs decay}

As the number of the produced gauge field quanta grows, they start
to affect the dynamics of the oscillating higgs field and eventually
shut down the resonance. The subsequent evolution towards thermal
equilibrium is a highly complicated process which we do not aim to
address here. Because of the non-linear dynamics of the later stages
of the resonance, it is also a non-trivial task to precisely
determine the time when the higgs condensate has decayed. For our
purposes it suffices to estimate the decay time by the time when the
induced higgs mass $m_{h(W,Z)}^{2}$ generated by the excited gauge
bosons becomes equal to the mass term $m_{h}^{2}=3\lambda h^{2}$. At
this stage the backreaction becomes significant and our analysis can
no longer be applied to describe the system.

The higgs coupling to weak gauge bosons in the unitarity gauge takes
at large field values $h \gg v$ the form
\beq
\label{Lint}
\mathcal{L}_{{\rm int}}=q_{W}\lambda h^{2}W^{\mu+}W_{\mu}^{-}+\frac{1}{2}q_{Z}\lambda h^{2}Z^{\mu}Z_{\mu},
\eeq
where $q_{W}=g^2/\left(4 \lambda\right)$ and
$q_{Z}=\left(g^2+g'{}^2\right)/\left(4 \lambda\right)$ as before.
The effective higgs mass generated by the $W_{\mu}^{\pm}$ bosons is
then given by
\beq
\label{hWmass}
m_{h(W)}^{2}=2q_{W}\lambda\left\langle
W^{\mu+}W_{\mu}^{-}\right\rangle =2q_{W}\lambda\left\langle
W^{2}\right\rangle\ ,
\eeq
and the mass generated by $Z_{\mu}$ is obtained by replacing
$2q_{W}$ by $q_{Z}$. The resonantly produced higgs particles also contribute to the effective mass of the condensate but this is a
subdominant effect as the higgs quanta are produced at a much lower
rate than gauge bosons.

The expectation value in equation (\ref{hWmass}) can be approximated
as \cite{StructureOfResonance}
\beq
\label{W2}
\left\langle W^{2}\right\rangle \approx
\frac{1}{\left(2\pi\right)^{3}}\int\frac{d^{3}k\,n_{k}}{\sqrt{k^{2}+q_{W}\lambda
h_{\rm osc}^{2}}}\ ,
\eeq
where we have normalized the scale factor to unity at the onset of
higgs oscillations $a_{\rm osc}=1$. Only gauge bosons with rather
low momenta are produced, so we can approximate $k^{2}+q_{W}\lambda
h_{\rm osc}^{2}\approx q_{W}\lambda h_{\rm osc}^{2}$. For the number
density of produced particles we use the fitting $n_{k}\propto
\exp\left(4.6\mu_{k}z^{1/3}\right)$. To obtain a crude estimate, we
approximate the occupation numbers by $n_{k} \approx n_{k=0}$ up to
the cutoff scale $k_{*}\approx
\left(\frac{q_{W}}{2\pi^{2}}\right)^{1/4} \sqrt{\lambda}h_{\rm osc}$
\cite{StructureOfResonance} and set $n_{k}=0$ above this scale. This
yields the estimate
\beq
\left\langle W^{2}\right\rangle
\approx\frac{n_{k}}{\left(2\pi\right)^{3}\sqrt{q_{W}\lambda h_{{\rm
osc}}^{2}}}\frac{4\pi}{3}k_{*}^{3}\approx1.8\cdot10^{-3}q_{W}^{1/4}\lambda
h_{{\rm osc}}^{2}n_{k}\ .
\eeq

We approximate the time scale of the higgs decay by the time when
the backreaction of the decay products becomes significant
$m_{h(W)}^{2}\approx m_{h}^{2}$, which occurs at
\beq
\label{zdec}
n_{k}\sim\frac{1.3\cdot 10^{3}}{2q_{W}^{5/4}\lambda z_{\rm dec}^{2/3}}\ ,
\eeq
where we have make use of the fact that the
scale factor in the matter dominated background is given by
$a\approx (3z/2)^{2/3}$ for $z\gg 1$. From Eq.~(\ref{zdec}) with $n_{k}\propto
\exp\left(4.6\mu_{k}z^{1/3}\right)$ the time of higgs decay $z_{\rm dec}$ can be solved. As
$t_{\rm dec}= z_{\rm dec} H_{\rm osc}^{-1}$ for $t\gg t_{\rm osc}$,
the value $z_{\rm dec}$ directly gives the number of Hubble times
from the onset of higgs oscillations to the decay of the higgs
condensate. In (\ref{zdec}) we have for simplicity assumed the
production of $W$ gauge bosons dominates over the $Z$ boson
production. If the production of $Z$ gauge bosons is dominant
$q_{W}^{5/4}$ in (\ref{zdec}) should be replaced by $q_{Z}^{5/4}/2$.

\begin{table}[ht]
\caption{The amount of Hubble times $H_{\rm osc}/H_{\rm dec}$ from
the onset of higgs oscillations to the higgs decay. The number of
inflaton oscillations $n^{\rm dec}_{\phi}$ from the end of inflation
until the higgs decay is also shown.} \centering
\begin{tabular}{c c c c}
\hline\hline
$H_*$/GeV & $\lambda$ & $H_{\rm osc}/H_{\rm dec}$ & $n^{\rm dec}_{\phi}$ \\ [0.5ex]
\hline
$10^4$ & $0.09$ & $370$ & $1\,000$ \\
$10^6$ & $0.04$ & $360$ & $1\,700$ \\
$10^8$ & $0.02$ & $630$ & $5\,100$ \\
$10^{10}$ & $0.005$ & $340$ & $7\,700$ \\ [1ex]
\hline
\end{tabular}
\label{table:HubbleTime}
\end{table}
While equation (\ref{zdec}) admits a general analytical solution, we
will here only list its solutions for a set of parameter values. The
results are shown Table in (\ref{table:HubbleTime}) which shows that
the higgs condensate typically decays in a time scale of a few
hundred of Hubble times after the onset of oscillations, $H_{\rm
osc}/H_{\rm dec} = {\cal O}(10^2)$. We have also listed the
corresponding number of inflaton oscillations measured from the end
of inflation
\beq
n_{\phi}^{\rm dec} = \frac{m_{\phi} t_{\rm dec} }{2\pi} \sim
\frac{1}{2\pi}\frac{t_{\rm dec}}{H_*^{-1}}\ .
\eeq
Comparing with the results of Sect.~3, we find that the
non-perturbative decay occurs way before the perturbative decay
channels would become efficient. Looking at the cases shown in Table
(\ref{table:HubbleTime}) we observe that decay is slowest for
$H_*=10^8$GeV for which one of the resonance parameters is outside
the resonance bands and the other one happens to be on the edge of a
resonance band. Even in this case the higgs decay through the
non-perturbative channels occurs much before the perturbative
channels become relevant.

\subsection{Resonant production of quarks and leptons}

The Standard Model higgs ($h \gg v$) couples to quarks and leptons
in the unitarity gauge according to
\beq
\mathcal{L}_{{\rm int}}=-\sqrt{q_{i}\lambda}hf\bar{f},
\eeq
where $q_{i}=\frac{m_{i}^{2}}{\lambda
h^{2}}=\frac{y_{i}^{2}}{2\lambda}$. As the oscillating higgs field
crosses zero fermions can be resonantly produced. Due to the fermion
statistics, the generated fermions can populate states only up to
the fermisphere whose radius is determined by the width of the
resonance and grows along with the expansion of space
\cite{PreheatingOfFermions}. The number density of fermions
generated by the higgs decay therefore cannot compete with the
exponentially growing number of gauge bosons produced by the bosonic
resonance channels.

Since $\mathcal{L}_{{\rm int}}\propto h$ the fermions do not give
rise to effective mass for the higgs condensate at tree level
contrary to the gauge bosons. However, the fermions do change the
equation of motion of the higgs field (\ref{higgseom}) via
backreaction. In the Hartree approximation one finds
\beq
\frac{\ddot{h}}{h_{{\rm osc}}}+3\frac{H}{\sqrt{\lambda}h_{{\rm osc}}}\frac{\dot{h}}{h_{{\rm osc}}}+\left(\frac{h}{h_{{\rm osc}}}\right)^{3}+\frac{\sqrt{q_{i}}}{\sqrt{\lambda}h_{{\rm osc}}^{3}}\left\langle f\bar{f}\right\rangle =0
\eeq
from which one could examine the backreaction effects of the
generated fermions and the associated time scales, see
\cite{MassiveFermions}. However, as the number density of the
resonantly generated gauge bosons grows exponentially as opposed to
the fermions, it is clear that the dominant decay channels of the
higgs condensate are the bosonic ones. For our purposes, the
fermionic channels are therefore irrelevant and we will not address the
details of the resonant fermion production further.

\section{Discussion}

The remarkable detection of a new boson consistent with the Standard
Model higgs \cite{Aad:2012tfa} by the ATLAS and CMS collaborations
also suggest that during the inflationary epoch there was at least
one other scalar in addition to the inflaton field. The minimal
scenario where the higgs itself would be the inflaton
\cite{Bezrukov:2007ep} unfortunately requires a large non-minimal
coupling to gravity and therefore does not seem to represent the
generic situation.

In the present paper we have explored the fate of the minimally
coupled Standard Model higgs during and after inflation assuming the
Standard Model remains valid up to inflationary energy scales. The
effective higgs potential is then exactly calculable in terms of the
measured Standard Model parameters. We have shown that for any
initial higgs value in the range where the Standard Model vacuum is
stable, the expectation value of the higgs mass becomes much smaller
than the Hubble scale over a period of less than about 30 efolds of
inflation. The higgs then generically corresponds to a light
spectator field during inflation and it acquires a spectrum of
nearly scale invariant perturbations over the observable scales. The
distribution of the higgs field relaxes to an equilibrium form with
a calculable variance in between 20 and 85 efoldings, depending on
the inflationary scale $H_*$.

After the end of inflation the higgs becomes effectively massive and
starts to oscillate in less than ${\cal O}(10^2)$ Hubble times. We
find that the oscillating higgs condensate generically decays into
weak gauge bosons through a parametric resonance within a few
hundreds of Hubble times from the onset of oscillations. The precise
number depends on the inflationary scale $H_*$ as shown in Table
\ref{table:HubbleTime}.

At the time of inflation the higgs condensate contributes very
little to the total energy density and its perturbations amount to a
small isocurvature component. The higgs perturbations could however
be a source of significant metric perturbations at a later stage if
the expansion history after inflation is sensitive to the exact
value of the higgs condensate. This could happen for example through
the modulated reheating mechanism with the higgs modulating the
inflaton decay \cite{DeSimone:2012qr,DeSimone:2012gq,Choi:2012cp}.
Whether such a conversion takes place depends on details of the
inflaton sector beyond Standard Model which we have not specified
here. However, a general necessary condition for the scenario is
that the inflaton decay should occur before the decay of the higgs
condensate. As we have shown in this work, the decay of the Standard
Model higgs takes place after $n_{\phi} ={\cal O}(10^3)$
oscillations of the inflaton field. A necessary condition for
generating significant perturbations through a modulated reheating
scenario with the higgs therefore is that the inflaton couplings
should be large enough for it to decay within this time. Similar
remarks can be made on the curvaton scenario with the Standard Model
higgs although the viability of the mechanism is hampered by
the quartic form of the higgs potential \cite{Choi:2012cp}.

As the Standard Model higgs generically acquires perturbations
during inflation it would be interesting to specify the general
conditions under which its perturbations can significantly
contribute to the observable primordial perturbations. Our findings
suggest that curvaton-like conversion mechanisms, where the inflaton
has to decay before the higgs, imply relatively strong couplings on
the inflaton sector to be efficient. If the inflaton decays directly
into Standard Model degrees of freedom it would be interesting to
investigate if the required strength of the couplings is consistent
with assuming negligible modifications to the effective potential
for Standard Model fields from the inflaton sector. We plan to
address these topics more carefully in a future work.

\acknowledgments{KE is supported by the Academy of Finland grants
1263714 and 1218322. TM is supported by the Magnus Ehrnrooth
foundation. SN is supported by the Academy of Finland grant 257532.}

\end{document}